\documentclass[a4paper]{article}

\usepackage{icrc2013}
\usepackage[english]{babel} 

\title{Study of the Crab Nebula TeV emission variability during five years with ARGO-YBJ}

\shorttitle{ARGO-YBJ Crab Nebula variability}

\authors{
S.Vernetto$^{1,2}$
for the ARGO-YBJ Collaboration.
}

\afiliations{
$^1$ Osservatorio Astrofisico di Torino - INAF - Italy \\
$^2$ Istituto Nazionale di Fisica Nucleare - Sezione di Torino - Italy \\

}

\email{vernetto@to.infn.it}

\abstract{The flaring activity of the Crab Nebula is one of the most puzzling 
phenomena of the gamma ray sky. The light curves 
in the energy range E $>$100 MeV show a high flux variability 
on time scales ranging from hours 
to weeks, with sharp emission peaks superimposed to long lasting 
smoother modulations, 
whose origin is still under debate. 
A long term observation of the Crab Nebula at 
TeV energies could add useful information to 
understand the mechanisms responsible of this unexpected behavior.
The air shower detector ARGO-YBJ 
monitored the Crab Nebula in the energy range 
0.5-20 TeV from November 2007 to February 2013.
During the flaring episodes observed by Fermi, 
the average ARGO-YBJ flux is found to be 
a factor 2.4 $\pm$0.8 larger than the average value.
Performing a long term study of the Crab Nebula flux, the ARGO-YBJ light curve  
is consistent with a uniform flux with a probability of 0.11.
However, a comparison with the Fermi LAT light curve during 4.5 years shows 
a correlation between the data of the two experiments.
The percent flux variations observed by ARGO-YBJ 
with respect to the average value are consistent with the
variations of the Fermi rate, suggesting, in the hypotesis that the
modulations are real, the same physical process
at the origin of the observed flux variations.

}


\begin{document}
\maketitle

%---------------------------------------------------------------------
\section{Introduction}

The Crab Nebula is the remnant of a supernova exploded in 1054 A.D.,
containing a pulsar that powers a wind of relativistic particles.
The interactions of these particles with the remnant gas and magnetic field
produce a radiation extending from radio waves to gamma rays.
 
Most of the emission is attributed to synchrotron radiation of relativistic
electrons and positrons. The spectral energy distribution (SED) peaks
between optical and X-ray frequencies. A second component arises above 400 MeV,
interpreted as Inverse Compton of the same electrons scattering off synchrotron photons
and CMB photons. 

The large gamma ray flux and its assumed stability, made the Crab Nebula
a ``standard candle'' for gamma ray astronomy, suitable to calibrate instruments.
Unexpectedly, on 2010 September the AGILE satellite detected a strong flare
from the direction of the Crab Nebula at energies above 100 MeV, lasting 2 days, 
with a maximum flux 3 times larger
than the average value \cite{bib:agile1}, 
later confirmed by Fermi \cite{bib:fermi1}. 
From then on, Fermi and AGILE reported a few more flares, characterized by
a rapid increase and decay of the flux, tipically lasting a few days.
The most impressive occurred on 2011 April, when the observed 
flux was $\sim$10 times larger than usual \cite{bib:fermi2}.
Besides flares, the Crab Nebula shows smaller flux variations 
on longer time scales, called "waves" in \cite{bib:striani}.

The origin of all these events is still under debate. 
The flux variations are attributed to the Nebula, since the Pulsar emission was found 
to be stable within 20$\%$ \cite{bib:fermi2}.
The measured SED shows a new spectral component emerging during flares, peaking at high energies
(up to hundreds MeVs in the 2011 April flare), attributed to a synchrotron emission
of a population of electrons accelerated up to energies of 10$^{15}$ eV.
The site and the origin of such a surprising activity is unknown.

In this uncertain scenario, 
observations at higher energies could provide useful 
information to shed light on the puzzle.

The ARGO-YBJ experiment is an air shower detector
located at Yangbajing (Tibet, China) 
at an altitude of 4300 m above the sea level, 
devoted to gamma ray and cosmic ray studies in the TeV energy range.
Due to the high duty cycle and the large field of view ($\sim$ 2 sr) 
ARGO-YBJ can observe every day a large part of the sky, monitoring
the flux of the most luminous gamma ray sources 
\cite{bib:m501,bib:cyg,bib:mgro,bib:m421b,bib:m421a}

A preliminary analysis of the data recorded by ARGO-YBJ during the flares 
occurences,
showed an increase of the Crab flux by a factor 4-5 with a moderate 
statistical significance, in 3 out of 4 flares \cite{bib:ver_scineghe}.

These observations have not been confirmed by Cherenkov telescopes, that could not monitor the
flaring episodes because of the presence of the Moon in the night sky.
Sporadic and short time measurements performed during the first part 
of the 2010 September flare show no evidence for a flux
variability \cite{bib:crab_cheren1,bib:crab_cheren2}. 

In this work we present a reanalysis of the ARGO-YBJ data, focalizing 
the study non only 
during the flaring days, but on the whole observation time of the 
Crab Nebula (five years).
The results of a correlation with Fermi data 
over 4.5 years are also reported.

%-----------------------------------------------------------------------------

\section{The ARGO-YBJ experiment}

ARGO-YBJ is an full coverage detector consisting of a 
$\sim$74$\times$ 78 m$^2$ carpet made of a single layer of Resistive
Plate Chambers (RPCs) with $\sim$93$\%$ of active area, sorrounded
by a partially instrumented ($\sim$20$\%$) area up to
$\sim$100$\times$110 m$^2$. 
The apparatus has a modular structure,
the basic data acquisition element being a cluster (5.7$\times$7.6
m$^2$), made of 12 RPCs (2.85$\times$1.23 m$^2$). 
Each RPC is read by 80 strips of 6.75$\times$61.8 cm$^2$ (the
spatial pixels), logically organized in 10 independent pads of
55.6$\times$61.8 cm$^2$ which are individually acquired and
represent the time pixels of the detector.  
The full experiment is made of 18360 pads 
for a total active surface of $\sim$6600m$^2$. 
 
The showers firing a number of pads N$_{pad} \geq$20 in the central carpet 
generate the trigger.
The time of each fired pad and its location are recorded
and used to reconstruct the position of the shower
core and the arrival direction of the primary particle.

The angular resolution and the pointing accuracy of the detector
have been evaluated by using the Moon shadow, observed
by ARGO-YBJ with a statistical significance 
of $\sim$9 standard deviations per month.
The shape of the shadow
provides a measurement of the point spread function
(PSF), which is in eccellent agreement with a Monte Carlo simulation \cite{bib:DiS11}. 
The simulated angular resolution for gamma rays
is smaller by $\sim$30-40$\%$ with respect to the
angular resolution for cosmic rays, due
to the better defined time profile of the showers. 

The Moon shadow has also been used to check the absolute energy calibration
of the detector, by studying the westward shift of the shadow
due to the geomagnetic field.
From this analysis the total absolute energy scale error,
including systematics effects, is estimated to be less than 13$\%$ 
\cite{bib:DiS11}.

The full detector has been in stable data taking 
from 2007 November to 2013 February with a duty cycle $\sim
86\%$. The trigger rate is $\sim$3.5 kHz with a dead time of 4$\%$.

%--------------------------------------------------------------------

\section{Data analysis}

At the ARGO-YBJ site, the Crab Nebula culminates with a
zenith angle of 8$^{\circ}$ and every day is visible for 5.8 hours 
with a zenith angle less than 40$^{\circ}$.
The dataset used in this analysis contains all the events recorded
from November 2007 to February 2013, with N$_{pad} \geq$20,
where  N$_{pad}$ is the number of hit pads on the central carpet.
The total on-source time is 9520 hours. 

For each source transit, 
the events are used to fill a set of 16$^{\circ}\times$16$^{\circ}$ 
sky maps in celestial coordinates (right ascension and declination) with
0.1$^{\circ}\times$0.1$^{\circ}$ bin size, centered on the Crab Nebula
position, each map corresponding to a defined N$_{pad}$ interval.
We use 8 intervals, corresponding to  N$_{pad}$=20-39, 40-59, 60-99, 100-199,
200-299, 300-499, 500-999 and  N$_{pad}>$1000.

In this new analysis, the arrival directions of the showers have been corrected
for a systematic inclination that affects the events with the core falling 
far from the detector center, according to the method described in \cite{bib:eck}.
The improved angular resolution increases the sensitivity
by a factor 1.1, 1.3 and 
1.9 for events with N$_{pad}\geq$40, 300 and 1000, respectively.
For the same class of events, the radius of the opening angle 
that optimizes the signal-to-background ratio 
is 0.97$^{\circ}$, 0.39$^{\circ}$ and 0.30$^{\circ}$, respectively,
and contains 46$\%$, 58$\%$, and 64$\%$ of the signal.

% \begin{figure}[t]
%  \centering
%  \includegraphics[width=0.4\textwidth]{paper_gauss.eps}
%  \caption{The Crab Nebula spectrum obtained by ARGO-YBJ, compared with other
%measurements \cite{bib:hess,bib:magic}.}
%  \label{spe}
% \end{figure}

In order to extract the excess of $\gamma$ rays, the cosmic ray
background is estimated with the {\em time swapping} method
and it is used to build the ``background maps'' \cite{bib:ale92}.
%For each detected
%event, $n$ "fake" events (with $n$ = 10) are generated by replacing the original
%arrival time with new ones, randomly selected from an event buffer that
%spans a time T of data taking. 
%Changing the time, the fake events maintain the same declination of 
%the original event,
%but have a different right ascension. With these events a new sky
%map (background map) is built, with a statistics $n$ times larger
%than the ``true'' event map in order to reduce the fluctuations. 
%To avoid the inclusion of the source events in the background evaluation,
%the showers inside a circular region around the source (with a radius 
%related to the PSF and depending on N$_{pad}$) are excluded 
%from the time swapping procedure. A correction of the number of swaps 
%is made to take into account 
%the rejected events in the source region.
%The value of the swapping time T is $\sim$ 3 hours, in order to minimize the
%systematic effects due to the environmental parameters variations.

The maps are then smoothed according to the corresponding PSF.
Finally, the smoothed background map is subtracted to the
smoothed event map, obtaining the "excess map",
where for every bin the statistical significance of the excess is
evaluated. For a detailed account of the analysis technique 
see \cite{bib:mgro}.

%given by:

%\begin{displaymath}
%S = (N_{on}-N_{off})/\sqrt{\delta N_{on}^2+ \delta N_{off}^2)} 
%\end{displaymath}

%with $N_{on}$ = $\Sigma_i$ $N_i$ $w_i$  and
%$N_{off}$ = $\Sigma_i$ $B_i$ $w_i$ /$n$. 
%In these expressions $N_i$ and $B_i$  are the number of events 
%of the i$^{th}$ bin of the ``event map'' and ``background map''
%respectively,  
%$w_i$ is a weight, proportional to the value of the PSF 
%at the angular distance of the i$^{th}$ bin, and
%$n$ is the number of swappings.
%The sum is over all the bins inside a radius R, chosen 
%to contain the PSF.
%Since the number of events per bin is large, 
%the fluctuations follow the Gaussian
%statistics, hence the errors on $N_{on}$ and $N_{off}$ are:
%    $\delta N_{on}$ =  $\sqrt{\Sigma_i N_i w_i^2}$
%and $\delta N_{off}$ = $\sqrt{\Sigma_i B_i w_i^2/n^2}$.

Adding all the transits,
an excess at the source position is observed in every map, 
with a significance of
4.1, 7.4, 8.7, 7.0, 7.5, 6.7, 4.4 and 5.2 standard deviations, respectively,
for a total significance of 19 standard deviations.

The average spectrum is evaluated by means of a simulation,
by comparing the number of excess events for each N$_{pad}$ interval,
with the corresponding values expected assuming a set of test spectra.
Assuming a power law spectrum, the obtained best fit 
in the energy range $\sim$0.5-20 TeV is:
dN/dE = 4.6$\pm$0.20 $\times$ 10$^{-12}$(E/2 TeV)$^{-2.67\pm0.05}$ 
photons cm$^{-2}$ s$^{-1}$ TeV$^{-1}$,
with a $\chi^2$ of 8.1 for 6 degrees of freedom.

%The quoted errors are purely statistical.
%We evaluate a systematic error on the flux less than 30$\%$ mainly due 
%to the background 
%evaluation and to the uncertainty on the absolute energy scale.

%Fig.1 shows the obtained spectrum, 
%in agreement with previous
%measurements by other detectors in the same energy range. 
%The energy of each flux point
%represents the gamma ray median energy in the corresponding  N$_{pad}$ interval.

In order to study the time behaviour of the Crab Nebula emission,
we used the events with N$_{pad}\ge$40 to
avoid threshold effects. All the source transits with an observation time
less than 0.5 hours have been discarded from the analysis. 
The total number of useful days is 1816.

Fig.1 shows the distribution of $\sigma _i$=($R_i-R_m$)/$\delta (R_i-R_m)$, 
where $R_i$ is the Crab counting rate of the i-th day,
$R_m$ is the average counting rate ($R_m$=18.5$\pm$1.4 ev h$^{-1}$), and
$\delta (R_i-R_m)$ is the statistical error on the difference $R_i-R_m$.
The daily rates are corrected for the detector efficiency and the
variations of atmospheric conditions (see Subsection 4.1).

The distribution can be fitted by a Gauss function with a mean value 
0.001$\pm$0.028 and r.m.s.= 1.074$\pm$0.019.
The value of the r.m.s. 
indicates variations slighty larger than the expected statistical fluctuations.

 \begin{figure}[t]
  \centering
  \includegraphics[width=0.45\textwidth]{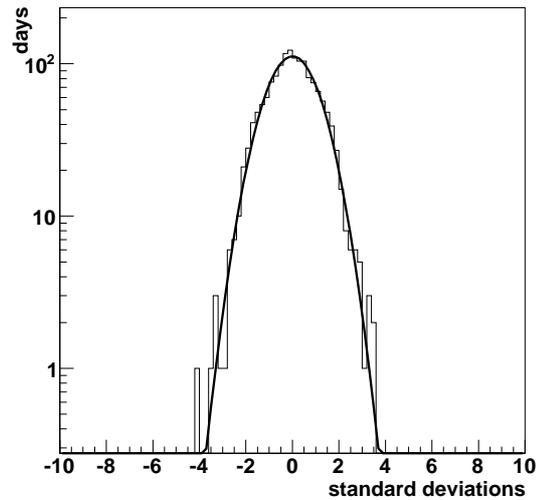}
  \caption{Distribution of the daily excesses from the Crab Nebula around 
the average value, in units of standard deviations, measured by ARGO-YBJ.}
  \label{gauss}
 \end{figure}

%---------------------------------------------------------------------------

\section{Correlation with Fermi data}

To study possible time correlations between ARGO-YBJ and Fermi, 
we used the 
Fermi LAT daily light curve from August 2008 to February 2013 
for energy $E>$100 MeV, obtained by the authors 
of \cite{bib:striani} with a standard unbinned likelihood analysis.

A Fourier spectrum of the light curve reveals a residual periodicity of 53.5
days with a semi-amplitude $\sim$7$\%$, likely due 
to the LAT instrument precession \cite{bib:fermi_prec}.
This effect can be easily corrected since the oscillation is well fitted
by a sinusoidal function. Fig.2 shows the corrected light curve, representing
the sum of the Nebula and Pulsar fluxes.
The average flux is (2.66$\pm$0.01)$\times$ 10$^{-6}$ ph cm$^{-2}$ s$^{-1}$.
Even excluding the days with flares, the rate
is cleary variable, with modulations on time scales of weeks and months.

Since the ARGO-YBJ sensitivity does not allow to observe a significant signal
during a single flare, the Fermi data have been grouped according to the
measured flux, and for any group the average ARGO-YBJ flux is evaluated.
We have selected the days in which the Fermi flux is larger than 
$F_{min}$, with $F_{min}$ ranging from 2.5 $\times$ 10$^{-6}$ to 5.0 $\times$ 10$^{-6}$ 
ph cm$^{-2}$ s$^{-1}$. 
For any value of $F_{min}$, Table 1 shows the number of days
satisfying this condition, and the corresponding ARGO-YBJ flux, 
averaged over the same days.

According to the data, the ARGO-YBJ rate increases with the the Fermi flux,
indicating that there could be some
correlation between the flux measured by the two
detectors. During flares (here defined as the days in which the Fermi 
flux exceeds 5 $\times$ 10$^{-6}$ ph cm$^{-2}$ s$^{-1}$), the ARGO-YBJ
flux is a factor 2.4$\pm$0.8 larger than the average value.

\begin{table}[h]
\begin{center}
\begin{tabular}{|c|c|c|}
\hline Fermi flux & Number of days & ARGO-YBJ rate\\ 
       (ph cm$^{-2}$ s$^{-1}$) &   & (ph h$^{-1}$)\\ \hline
$>$2.5 $\times$ 10$^{-6}$  & 996 & 18.5$\pm$1.9   \\ \hline
$>$3.0 $\times$ 10$^{-6}$  & 470 & 19.8$\pm$2.8   \\ \hline
$>$3.5 $\times$ 10$^{-6}$  & 162 & 23.2$\pm$4.7   \\ \hline
$>$4.0 $\times$ 10$^{-6}$  & 56  & 32.8$\pm$8.0   \\ \hline
$>$4.5 $\times$ 10$^{-6}$  & 27  & 34.6$\pm$11.4  \\ \hline
$>$5.0 $\times$ 10$^{-6}$  & 17  & 44.7$\pm$14.3  \\ \hline
\end{tabular}
\caption{The Crab Nebula rate measured by ARGO-YBJ, for different levels of the flux measured by Fermi.}
\label{table1}
\end{center}
\end{table}

 \begin{figure}[!t]
  \centering
  \includegraphics[width=0.45\textwidth]{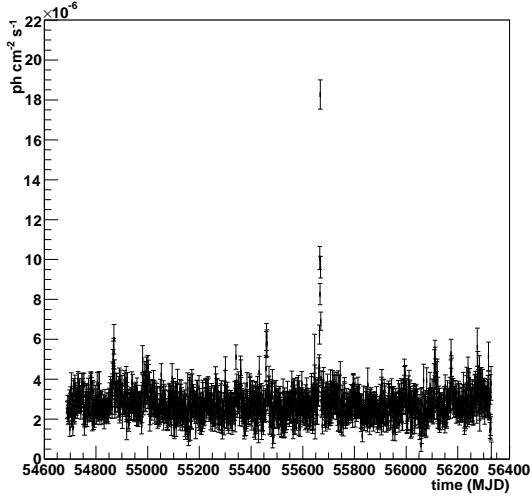}
  \caption{Crab (Nebula + Pulsar) daily light curve measured by Fermi.}
  \label{fermi_lc}
 \end{figure}

 \begin{figure}[]
  \centering
  \includegraphics[width=0.45\textwidth]{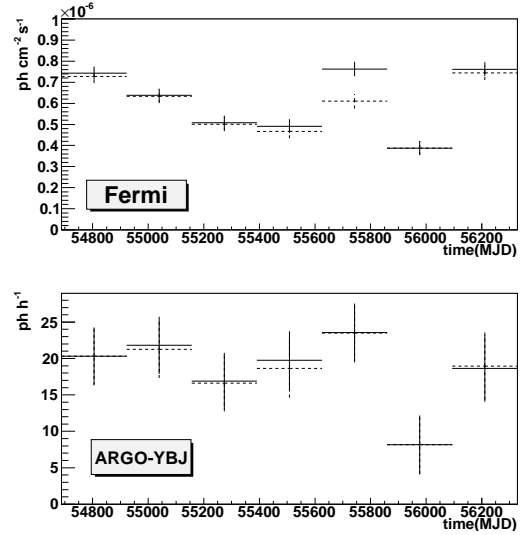}
  \caption{Upper panel: Crab Nebula flux measured by Fermi. 
           Lower panel: Crab Nebula flux measured by ARGO-YBJ. 
           The dashed line in both panels 
           has been obtained excluding the days with flares.}
  \label{correl}
 \end{figure}

An other way to search for a time correlation is comparing
the light curves of the two experiments.
We rebuild the Fermi light curve using a bin width larger than 
200 days in order to have a significant signal in the ARGO-YBJ data. 
We choose a bin width of 234 days
in order to divide exactly the period from MJD 54688 (start of the Fermi data)
to MJD 56328 (stop of the ARGO-YBJ data) in 7 bins.
  
Since the flux measured by Fermi is the sum of the Nebula 
and the Pulsar contributions, and since
the Pulsar flux $F_P$ averaged over the pulsation period is stable
($F_P$ = (2.04$\pm$0.01) $\times$ 10$^{-6}$ ph cm$^{-2}$ s$^{-1}$
for E $>$100 MeV \cite{bib:fermi2}),
the flux of the Pulsar has been subtracted.
The obtained Nebula flux is showed in the upper panel of Fig.3.
The flux shows significant variations, 
up to $\sim$40$\%$ of the average flux 
$f_N$ = (6.15$\pm$0.12) $\times$ 10$^{-6}$ ph cm$^{-2}$ s$^{-1}$.
The $\chi^2$ value is 127 for 6 degrees of freedom.
The large variations are not only due to flares.
In the same figure the dashed curve shows the flux
obtained excluding the flaring days, i.e. 17 days in which the total Crab
flux exceeds 5 $\times$ 10$^{-6}$ ph cm$^{-2}$ s$^{-1}$  ($\chi^2$=101).

The lower panel of Fig.3 shows the corresponding ARGO-YBJ data. 
The rate mean value for this period is 18.3 $\pm$ 1.5 ev h$^{-1}$.
The $\chi^2$ value is 11.7. 
Discarding the flaring days, $\chi^2$=11.2 (dashed curve).
Even if the ARGO-YBJ rate variations are not inconsistent 
with statistical fluctuations (the probability of a constant flux is 0.11),
the Fermi and ARGO-YBJ data seems to follow a similar trend.
 
Fig.4 shows the ARGO-YBJ percent rate variation  
with respect to the mean value ($\Delta F_{ARGO}$)
as function of the corresponding variation
of the Fermi rate  ($\Delta F_{Fermi}$), for the 8 bins of the light curve.

The data appear linearly correlated.
The Pearson correlation coefficient between the ARGO-YBJ and FERMI data 
is $r$=0.76. 

Doing the same analysis with higher N$_{pad}$ thresholds
(N$_{pad}$ $>$ 100 and higher) no correlation is visible. 
The lack of correlation for more energetic events could be due the 
smaller statistics and to the larger fluctuations which could
hidden the effect (the rate of events with N$_{pad}$ $>$ 
100 is 3.2$\pm$0.4 ev h$^{-1}$).

Fitting the 7 points with the function
$\Delta F_{ARGO}$= $a$ $\Delta F_{Fermi}$ + $b$,
the values of the best fit parameters 
are $a$ = 0.85$\pm$0.36 and $b$ = 0.009$\pm$0.083, 
with $\chi^2$=3.5 for 5 d.o.f.
Discarding the flaring days, the value of the best fit parameters
are $a$ = 0.85$\pm$0.40 and $b$ = 0.008$\pm$0.084, 
with $\chi^2$=4.5.
The regression coefficient $a\sim$1 implies the same percent variation
in Fermi and ARGO-YBJ rates.

%-------------------------------------------------------------------------
\vspace*{0.5cm}

\subsection{Stability of the ARGO-YBJ data}

Concerning the ARGO-YBJ data, the possible causes of artificial
rate variations have been accurately studied
in order to exclude systematic effects.

Since the measured number of events from the source $S=N-B_{ts}$ 
is the difference between the number of events
$N$ detected in the source observational window (that contains the 
source events $S$ plus the cosmic ray
background $B$) and the number of background events $B_{ts}$
estimated with the time swapping method, one must analyse separately the
stability of the different contributions.

1) A loss of signal events $S$ could be produced by variations
of the pointing accuracy.
Studying the Moon shadow month by month, we have verified that 
the pointing is stable within 0.1 degree. Given the moderate angular resolution
for events with N$_{pad} >$40, such a value could produce
fluctuations of the signal of less than 2$\%$.

2) Atmospheric pressure and temperature variations can affect
the detector efficiency, that can also be altered by some RPC 
not working properly.

3) The atmospheric conditions produce also changes in the shower rate
of the order of a few percent, due to the different condition in which
the showers propagate.

The two latter effects modify $S$, $B$ and $B_{ts}$ 
by about the same factor (neglecting the different behaviour
of cosmic ray and gamma ray showers, that in this contest 
is a second order effect). 
This allows the use of $B_{ts}$ to correct the Crab rate,
multiplying the Crab flux of the i-th bin of the light curve 
by the correction factor $f_c$=$B_m$/$B_i$, where $B_m$ is the 
average estimated background rate, and $B_i$ is the average estimated
background rate in that bin. 
The light curve of Fig.3 has been corrected according to this method,
with $f_c$ ranging from 0.92 to 1.08. 

4) A further possible systematics could be an incorrect evaluation
of the background $B_{ts}$. Since the value of the background is about 10$^3$
larger than the source signal, a small error in the
background estimation could produce a large change in the source flux.
We have tested the stability of the background using four fake sources 
located at a distance of 3 degrees from the Crab Nebula in different
directions.
If the background estimation was the origin of the observed flux variations,
the fakes source flux would be affected by similar modulations.
The light curve of the fakes source are consistent with uniform distributions
with mean values consistent with zero, without any 
trend similar to the one observed for the Crab Nebula signal.

 \begin{figure}[]
  \centering
  \includegraphics[width=0.45\textwidth]{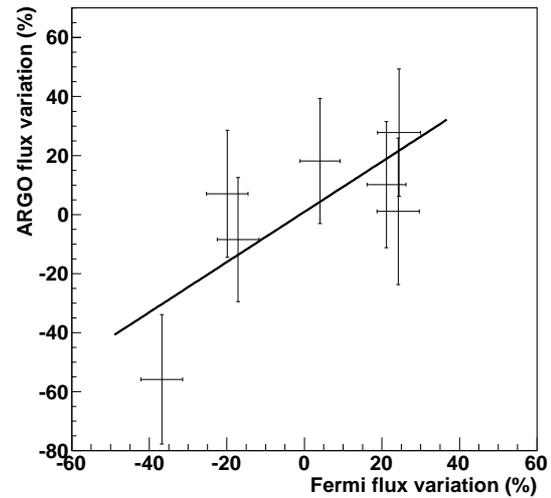}
  \caption{Percent variation of the Crab Nebula flux 
with respect to the average value: ARGO-YBJ vs Fermi data.}
  \label{correl2}
 \end{figure}

\section{Conclusions}

The ARGO-YBJ events recorded in 5 years have been reanalized
to study the time variability of the Crab Nebula emission in the TeV energy 
range. 

The ARGO-YBJ light curve 
is consistent with a constant flux with a probability of 0.11. 
However it shows modulations that appear 
correlated with the corresponding light curve
obtained with the Fermi LAT data.
According to our analysis, a percent variation 
of the Fermi flux corresponds to the
same percent variation of the ARGO-YBJ rate,
implying, in case of a real effect, a same behaviour
of the gamma ray emission at energies $\sim$ 100 MeV and $\sim$1 TeV.

Concerning the flaring periods, 
the Crab Nebula flux measured by ARGO-YBJ during the days in which
the Fermi flux exceeds 5 $\times$ 10$^{-6}$ ph cm$^{-2}$ s$^{-1}$
is a factor 2.4 $\pm$0.8 larger than the average value.

The small statistical significance of these results does not allow to claim the
detection of a flux variability and requires a confirmation
by more sensitive instruments.
So far, no variation of the Crab Nebula flux has been reported by any detector.
A TeV variable flux could hardly be interpreded 
as the Inverse Compton emission associated to the new synchrotron 
component observed during flares, and requires a new interpretation.
The linearity between the Fermi and the ARGO-YBJ fluxes, and the value of
the regression coefficient consistent with unity,
would suggest the same physical mechanism at the origin of the flux variations
observed at E$\sim$100 MeV and E$\sim$1 TeV.

\vspace*{0.5cm}
\footnotesize{{\bf Acknowledgment:}{ We acknowledge E.Striani and M.Tavani
for their kind collaboration in providing the Fermi light curve used in this
analysis.}


\begin{thebibliography}{}



\bibitem{bib:agile1} M.Tavani et al., Science 331 (2011) 736.
\bibitem{bib:fermi1} A.A.Abdo et al., Science 331 (2011) 739.
\bibitem{bib:fermi2} R.Buehler et al., ApJ 749 (2012) 26.
\bibitem{bib:striani} E.Striani et al., ApJ 765 (2013) 52
\bibitem{bib:m501} B.Bartoli et al., ApJ 758 (2012) 2.
\bibitem{bib:cyg} B.Bartoli et al., ApJL 745 (2012) 22.
\bibitem{bib:mgro} B.Bartoli et al., ApJ 760 (2012) 110.
\bibitem{bib:m421b} B.Bartoli et al., ApJ 734 (2011) 110.
\bibitem{bib:m421a} G.Aielli et al., ApJL 714 (2010) 208.
\bibitem{bib:ver_scineghe} S.Vernetto et al., Nucl.Phys.B 239-240C (2013) 98. 
\bibitem{bib:crab_cheren1} M.Mariotti et al., 2010, Astron.Telegram 2967.
\bibitem{bib:crab_cheren2} R.Ong et al., 2010, Astron.Telegram 2968.
\bibitem{bib:DiS11} B.Bartoli et al., Phys. Rev. D, 84 (2011) 022003.
\bibitem{bib:eck} R.Eckmann et al., Proc. 22nd ICRC, 1991, HE 3.6.16. 
\bibitem{bib:ale92} D.R. Alexandreas et al., NIM A328 (1993) 570.
\bibitem{bib:hess} F.A. Aharonian et al., A\&A, 457 (2006) 899.
\bibitem{bib:magic} J.Albert et al., ApJ 674 (2008) 1037.
\bibitem{bib:fermi_prec} http://fermi.gsfc.nasa.gov/ssc/data/analysis/

\end{thebibliography}
\end{document}